\documentclass[twocolumn]{aastex631}

\usepackage{amssymb,amsmath,verbatim,mathtools,needspace,enumitem,etoolbox,graphicx,physics,microtype,afterpage,xspace,tabularx,lmodern,multirow,orcidlink}

\shorttitle{Increased TDE rates at high-z}
\shortauthors{K.~Kritos \& J.~Silk}

\newcommand{\jhu}{\affiliation{William H. Miller III Department of Physics and Astronomy, Johns Hopkins University, 3400 North Charles
Street, Baltimore, MD 21218, USA}}

\newcommand{\iap}{\affiliation{Institut d’Astrophysique de Paris, UMR 7095 CNRS and UPMC, Sorbonne Universit\'e, F-75014 Paris, France}}

\newcommand{\bio}{\affiliation{Department of Physics, Beecroft Institute for Particle Astrophysics and Cosmology, University of Oxford, Oxford OX1 3RH, UK}}

\begin{document}

\title{Toward Black Hole Stars: supermassive black hole growth in nuclear clusters \\ via stellar-object and gas accretion}

\begin{abstract}
 Supermassive black hole (SMBH) growth plausibly occurs via runaway astrophysical black hole mergers in nuclear star clusters that form intermediate mass black hole seeds at high redshifts. Such a model 
 yields an order-of-magnitude higher rate of tidal disruption events than that of compact-object captures. Our prediction, normalized to our proposed resolution of SMBH seeding, yields detectable tidal disruption event rates at high redshift. The resulting dense gas cocoons generate compact galactic nuclei, each incorporating a central, massive, black hole star, with comparable masses in gas, stars, and massive black holes within a scale of around a parsec as inferred from the various Little Red Dot spectral signatures. \vspace{1cm}
\end{abstract}

\author{Konstantinos Kritos$\,$\orcidlink{0000-0002-0212-3472}}
\email{kkritos1@jhu.edu}
\jhu

\author{Joseph Silk$\,$\orcidlink{0000-0002-1566-8148}}
\jhu
\iap
\bio

\date{\today}

\section{Introduction}
\label{sec:Introduction}

Extreme nuclear transients (ENTs) are rare, ultraluminous flares ($>10^{45}\,\rm erg\, s^{-1}$), exceeding supernova energies and more easily detected at $z>1$ than classical tidal disruption events (TDEs) in flux-limited surveys. The JWST has also uncovered numerous “Little Red Dots” (LRDs; $4 < z < 8$), likely AGN powered by $\sim10^7$--$10^8\,M_\odot$ supermassive black holes~\citep[SMBHs;][]{Greene:2024phl,Maiolino:2023bpi,Matthee:2023utn}.

ENTs show slow decays ($\gtrsim150\,\rm days$), emit more total energy than TDEs, and have smooth ($<10\%$ variability) light curves, blue spectra, and broad lines distinguishing them from AGN variability or supernovae~\citep{Frederick:2020sjz}. They are proposed to be TDEs of high-mass stars ($>3\,M_\odot$) by SMBHs~\citep{2025SciA...11...74H}.

Typical TDE durations are $3\times10^6$--$10^7\,\rm sec$~\citep{Melchor:2025fey}. Rates are $\sim10^{-5}\,\rm galaxy^{-1}\, yr^{-1}$ in normal galaxies, enhanced by 10--100 times in poststarburst systems~\citep{vanVelzen:2020hrd}.
A new class of ambiguous nuclear transients (ANTs) is also observed~\citep{Wiseman:2024rkz}.

\cite{Bellovary:2025aeg} estimates a high-$z$ TDE rate of $\sim10^{-4}\,\rm yr^{-1}$ (for $5<z<8$) by linking LRDs to SMBH-seed densities assuming purely TDE-powered emission, a rate too low for SMBH growth and implying LRD masses below earlier claims. A $\sim10^7\,M_\odot$ black hole (BH) has meanwhile been directly measured in a lensed $z=7$ LRD~\citep{2025arXiv250821748J}. Since LRDs are likely AGN-dominated, \cite{Kritos:2025aqo} show that including SMBH-growth needs (mergers + accretion) significantly boosts high-$z$ TDE rates, which we argue here may be observable.

In this work, we analyze an $N$-body snapshot at time $t$ of a star cluster within a central SMBH’s sphere of influence by sampling its spherically symmetric distribution function without time evolution. From these snapshots, we compute the instantaneous TDE and compact-object capture rates, incorporating loss-cone physics and stellar evolution.

\section{Stellar feeding onto central hole}
\label{sec:Stellar_feeding_onto_central_hole}

We model a nuclear star cluster (NSC) around an SMBH of mass
$M_{\rm SMBH}$ and dimensionless spin parameter $a_{\rm SMBH}$. A star of mass $m_\star$ and radius $R_\star$
is removed if its pericenter lies within the loss-cone radius
$r_{\rm lc}=\max(r_{\rm T},r_{\rm mb})$, where
$r_{\rm T}=R_\star(M_{\rm SMBH}/m_\star)^{1/3}$~\citep{1989ApJ...346L..13E} and $r_{\rm mb}$ is the spin-dependent marginally
bound orbit. For $r_{\rm T}\gtrsim r_{\rm mb}$, the event produces an electromagnetic TDE~\citep{Mummery:2023meb}. Stars enter the loss cone via
relaxation and are removed on a crossing time.

We compute event rates with Monte Carlo simulations in the Keplerian potential
$\Phi=-GM_{\rm SMBH}/r$ within the influence radius $r_{\rm infl}$, containing
$N_\star\simeq2M_{\rm SMBH}/\overline{m}_\star$ stars. Stellar masses are drawn from
an initial mass function (IMF) and evolved with {\tt updated-BSE}~\citep{Banerjee:2019jjs}. Assuming
$n_\star\propto r^{-\gamma}$ we sample radii from $p(r)\propto r^{3-\gamma}$ and
orbits from the isotropic distribution $f(a,e)\propto a^{2-\gamma}e$~\citep[Eq.~(4.36)]{Merritt's-book}. Orbital periods follow from Kepler's law, $P\simeq2\pi GM_{\rm SMBH}m_\star(-E)^{-3/2}$.

Angular-momentum diffusion is driven by nonresonant relaxation (NRR) with a timescale~\citep[Eq.~(5.61)]{Merritt's-book}
\begin{equation}
t_{\rm NRR}\simeq\frac{1.8\,{\rm Gyr}}{\ln\Lambda}
\frac{10^7\,M_\odot\,{\rm pc^{-3}}}{\rho_\star}
\frac{1\,M_\odot}{m}
\left(\frac{\sigma_\star}{100\,{\rm km\,s^{-1}}}\right)^3,
\end{equation}
where $\sigma_\star=(1+\gamma)^{-1/2}(GM_{\rm SMBH}/r)^{1/2}$ and
$\ln\Lambda\simeq\ln N_\star$. In the Keplerian regime, we also include resonant
relaxation~\citep[RR;][]{1996NewA....1..149R,Gurkan:2007bj}, with an effective relaxation
time $t_{\rm R}=(t_{\rm NRR}^{-1}+t_{\rm RR}^{-1})^{-1}$~\citep[Eq.~11]{Hopman:2006qr}.

Feeding rates are computed using the loss-cone theory by \cite{1999MNRAS.306...35S}. Defining $\theta_{\rm D}=(P/t_{\rm R})^{1/2}$~\citep[Eq.~(12)]{1976MNRAS.176..633F}, the loss cone is empty for $\theta_{\rm lc}>\theta_{\rm D}$, with rate
$[\ln(2/\theta_{\rm lc})t_{\rm R}]^{-1}$, and full otherwise, with rate
$\theta_{\rm lc}^2/P$ per unit star. Summing over stars within $r_{\rm infl}$ yields the total instantaneous rate \(\Gamma_{\rm lc}(t|M_{\rm SMBH})\).

For compact objects, we additionally include capture by gravitational
bremsstrahlung~\citep[Eq.~(11)]{1989ApJ...343..725Q} and gravitational wave (GW)-driven inspirals, known as extreme mass-ratio inspirals (EMRIs). The
effective loss-cone radius is the maximum of tidal, bremsstrahlung, and marginally
bound radii. For EMRIs, $\theta_{\rm D}=(t_{\rm GW}/t_{\rm R})^{1/2}$ and the
full-loss-cone regime is absent~\citep{2003ApJ...590L..29A}. We compute $t_{\rm GW}$
using the fit of \citet{Mandel:2021fra}.

\section{Initial conditions}
\label{sec:Initial_conditions}

We adopt a canonical IMF spanning $0.08$--$150\,M_\odot$ at metallicity $Z=0.002$.
Stellar types from {\tt updated-BSE} are grouped into stellar classes indexed by A as ``main sequence (MS)'' (1--2), ``giants'' (3--6), ``cores''
(7--9), ``white dwarfs (WDs)'' (10--12), ``neutron stars (NSs)'' (13), and ``BHs'' (14).

Orbits are initialized with a thermal eccentricity distribution; departures from
thermal (e.g., superthermal) enhances captures and inspirals by increasing
the low-angular-momentum population.

We assume approximate energy equipartition within $r_{\rm infl}$, leading to mass
segregation~\citep{Hopman:2006xn}. Stars, WDs, and NSs follow density slopes
$\gamma_\star=1.5$, while stellar-mass BHs are more centrally concentrated with
$\gamma_{\rm BH}=2$.

Five representative time epochs ($10^0$, $10^1$, $10^2$, $10^3$, and $10^4\,\rm Myr$) and three SMBH masses ($10^5,\,10^6,\ {\rm and}\ 10^7\,M_\odot$) are considered with $a_{\rm SMBH}=0.5$.
We link $r_{\rm infl}=GM_{\rm SMBH}/\sigma^2$ to $M_{\rm SMBH}$ through the $M_{\rm SMBH}$–$\sigma$ relation~\citep[Fig.~3]{2011Natur.480..215M}. We assume the SMBH influence region is
fully populated ($M_\star\gtrsim M_{\rm SMBH}$); otherwise, rates would scale down
accordingly.

\section{Mass accretion rate}
\label{sec:Mass_accretion_rate}

\begin{figure}
    \centering
    \includegraphics[width=0.49\textwidth]{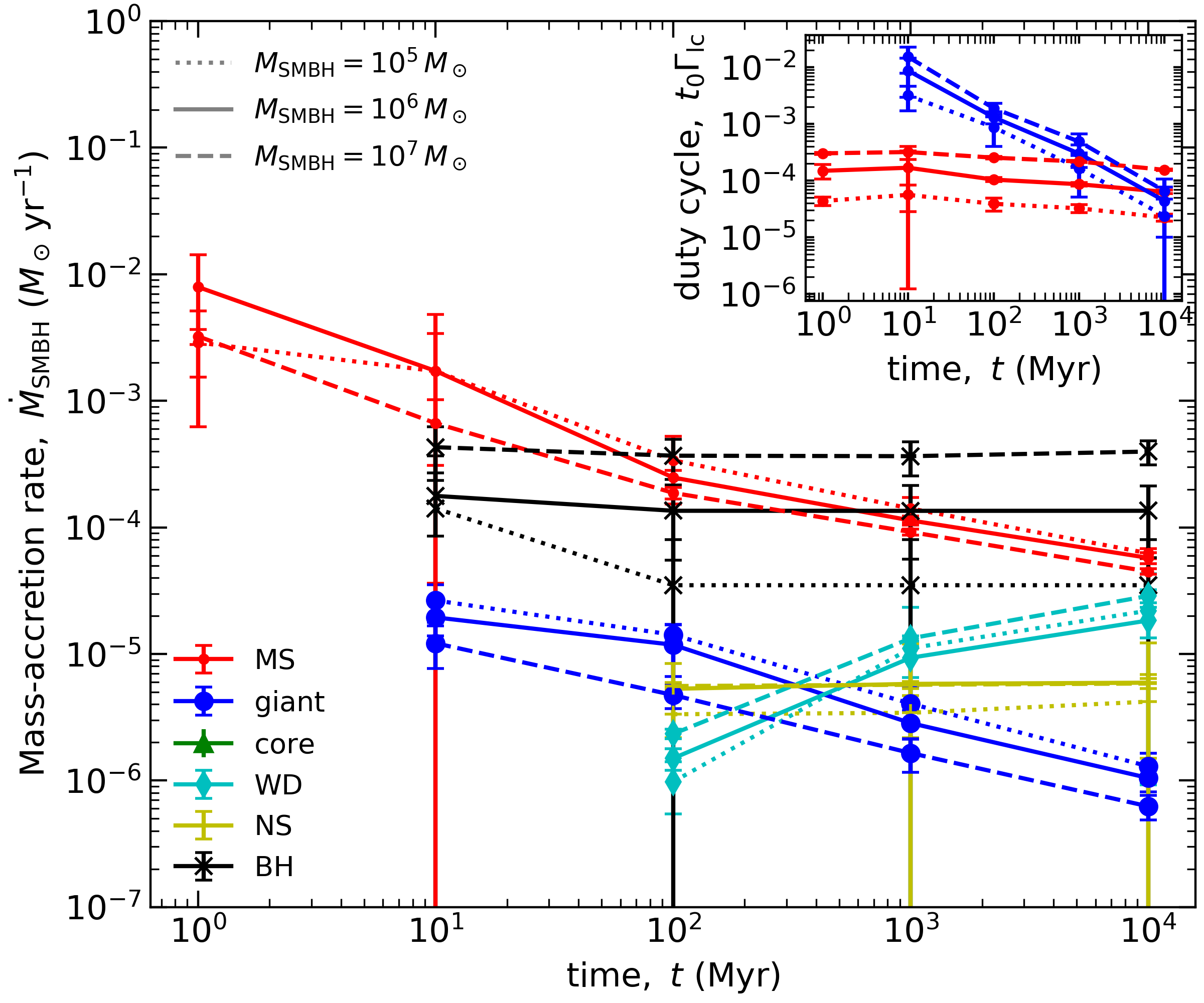}
    \caption{Temporal mass accretion rates from captures onto \(10^5\) (dotted), \(10^6\) (solid), and \(10^7\,M_\odot\) (dashed) SMBHs, decomposed by stellar type via loss-cone contributions. Abbreviations: MS (main sequence); WD (white dwarf); NS (neutron star); BH (black hole). Points with error bars indicate mean values and standard deviations from 10 realizations. The inset displays the duty cycle of MS (red) and giant (blue) stars.}
    \label{fig:massaccretionrate-time}
\end{figure}

Figure~\ref{fig:massaccretionrate-time} shows the SMBH mass accretion rate by stellar
type. Means and standard deviations are computed from 10 realizations per time bin.
Assuming a fraction $f_{\rm acc}$ of a stellar mass $m$ is accreted upon loss-cone
entry, the mass accretion rate per unit star is
$d\dot M_{\rm SMBH}^{(\rm A)}/dN_{\rm A}=f_{\rm acc}\,m\,d\Gamma_{\rm lc}^{\rm (A)}(t|M_{\rm SMBH})/dN_{\rm A}$ where $\rm A$ is the stellar class.
We adopt $f_{\rm acc}=0.5$ for MS stars (TDEs), $f_{\rm acc}=0.5$ for giants (envelope
disruption), and $f_{\rm acc}=1$ for compact remnants (BHs, NSs, WDs).

During the first $\sim100\,\rm Myr$, growth is dominated by MS stars, peaking at
$\sim10^{-2}\,M_\odot\,{\rm yr^{-1}}$ and initially driven by stars
$\gtrsim10\,M_\odot$. At later times ($t\gtrsim100\,\rm Myr$), massive stars evolve
into remnants and the disrupted MS population shifts to $\lesssim10\,M_\odot$.
Despite their larger loss cones, giants contribute $\sim2.5$ orders of magnitude
less mass than MS stars due to short lifetimes and limited envelope masses, becoming
negligible at late times.

MS and giant TDEs produce electromagnetic transients. The MS duty cycle
(remnant decay time $\times$ loss-cone rate) remains $\sim10^{-4}$, while giants
reach $\sim1\%$ owing to longer decay times, declining below the MS level after
several Gyr.

A few Myr after stellar-mass BH formation, the total accretion rate stabilizes at
$\sim10^{-4}\,M_\odot\,{\rm yr^{-1}}$ for a $10^6\,M_\odot$ SMBH and is higher by a
factor of a few for $10^7\,M_\odot$. Accretion is dominated by direct plunges, with
negligible inspirals. BHs dominate the late-time growth and contribute significantly even at early times due to mass segregation and their higher masses. WD accretion builds up over several Gyr to just below the MS contribution, while NSs maintain an
approximately constant rate of $\sim5\times10^{-6}\,M_\odot\,{\rm yr^{-1}}$ after
$\sim10$--$100\,\rm Myr$. Naked cores are negligible because of their short lifetimes and small radii.

Integrating $\dot M_{\rm SMBH}$ over $t\in[0,1]\,\rm Gyr$ yields total mass growths of
$\simeq3.9$, $0.45$, and $0.057$ times the initial SMBH mass for
$10^5$, $10^6$, and $10^7\,M_\odot$, respectively. The cumulative growth scales
approximately as $\propto\sqrt{t}$~\citep[also see][]{Stone:2016ryd}. These values represent
upper limits, as relaxation-driven ejections are neglected.

\section{Flaring events}
\label{sec:Flaring_events}

\begin{figure}
    \centering
    \includegraphics[width=0.49\textwidth]{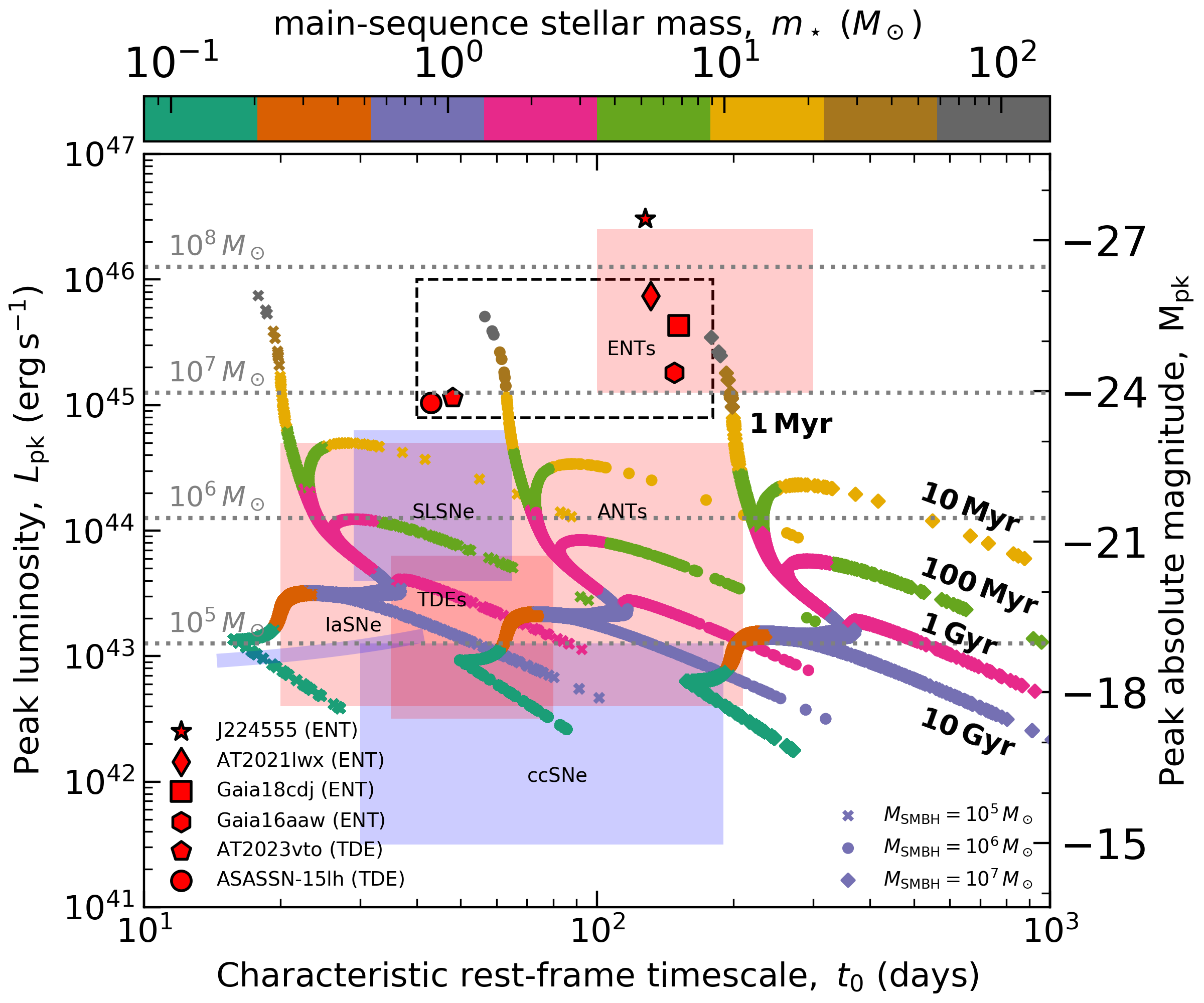}
    \caption{Peak luminosity versus characteristic rest-frame time-scale of astronomical transients, with peak absolute magnitudes normalized to a solar $V$-band magnitude of 4.83; ${\rm M}_{\rm pk}=2.5\log_{10}(L_{\rm pk}/L_\odot)-4.83$. Tidal disruption flares of MS stars by SMBHs of \(10^{5}\) (crosses), \(10^{6}\) (circles), and \(10^{7}\,M_\odot\) (diamonds) are shown as colored symbols, with the color bar indicating stellar mass. Five distinct sets correspond to five evolutionary times (bold labels). The six red symbols denote observed flares with \(L_{\rm pk}>10^{45}\,\rm erg\,s^{-1}\). Gray dotted horizontal lines mark the Eddington luminosities for the labeled SMBH masses. Abbreviations: ENTs (extreme nuclear transients); SLSNe (superluminous supernovae); ANTs (ambiguous nuclear transients); TDEs (tidal disruption events); ccSNe (core-collapse supernovae); Ia SNe (Type Ia supernovae) following the Phillips relationship.}
    \label{fig:luminosity-timescale}
\end{figure}

The SMBH disrupts MS stars, while giant-star TDEs are typically partial and involve
only envelope stripping \citep{Navarro:2024jxd}. The characteristic flare timescale
is the fallback time \citep[Eq.~(4)]{Gezari:2021bmb},
\begin{equation}
t_0 \simeq 40\,{\rm days}\,
\frac{1\,M_\odot}{m_\star}
\left(\frac{R_\star}{R_\odot}\right)^{3/2}
\left(\frac{M_{\rm SMBH}}{10^6\,M_\odot}\right)^{1/2},
\end{equation}
corresponding to the orbital period of the most bound debris. We adopt the
shock-powered emission model of \citet{Krolik:2024tvu}, in which stream
self-intersections near the apocenter produce radiation with efficiency
\citep[Eq.~(10)]{Krolik:2024tvu}
\begin{equation}
\eta \simeq 4.5\times10^{-4}
\left(\frac{M_{\rm SMBH}}{10^6\,M_\odot}\right)^{1/3}
\left(\frac{m_\star}{1\,M_\odot}\right)^{2/3}
\frac{R_\odot}{R_\star}.
\end{equation}
A complete predictive theory of TDE emission remains uncertain
\citep[e.g.,][]{Mummery:2025lgo}.

Owing to their large radii, giant-star TDEs produce longer, dimmer flares; we
therefore restrict our analysis to MS TDEs. Dense cores, naked cores, and WDs can
only be disrupted by intermediate-mass BHs \citep{Maguire:2020lad}.

In the efficient-cooling limit ($t_{\rm cool}\ll t_0$), the peak luminosity equals
the dissipation rate \citep[Eq.~(16)]{Krolik:2024tvu},
\begin{equation}
L_{\rm pk}\simeq 7.8\times10^{43}\,{\rm erg\,s^{-1}}
\left(\frac{\eta}{4.5\times10^{-4}}\right)
\frac{m_\star}{1\,M_\odot}
\frac{40\,{\rm d}}{t_0},
\end{equation}
well below $L_{\rm Edd}\simeq1.2\times10^{45}\,{\rm erg\,s^{-1}}
(M_{\rm SMBH}/10^6\,M_\odot)$. For inefficient cooling ($t_{\rm cool}\gg t_0$),
$L_{\rm pk}\sim L_{\rm Edd}$ and the flare duration is
$t_0\simeq m_\star c^2/L_{\rm pk}$ \citep[Eq.~(18)]{Krolik:2024tvu}.

Figure~\ref{fig:luminosity-timescale} shows the peak luminosities and rest-frame
decay times of MS TDEs in the efficient-cooling limit, colored by stellar mass,
alongside observed transient classes and six extreme flares
\citep{Graham:2025tla,2025SciA...11...74H}. The most luminous events arise from the
most massive MS stars. As the stellar population evolves, the maximum MS mass
decreases while stellar radii increase near turnoff, producing a late-time
turnover since $t_0\propto R_\star^{3/2}$. TDEs of $\sim30$--$150\,M_\odot$ MS stars
by $10^6$--$10^7\,M_\odot$ SMBHs are consistent with ENTs, while lower-mass systems
fall in the ANT regime.

\section{Integration over cosmological volume}
\label{sec:Integration_over_cosmological_volume}

We focus on the redshift range $z<6$. Assuming a
constant SMBH mass function with comoving number density $n_{\rm SMBH}$, we weight
three representative SMBH masses by $w_j\propto M_{{\rm SMBH},j}^{-2}$ and normalize
$\sum_j w_j=1$. Our choice of equal SMBH mass per unit logarithmic SMBH mass bin in the examined mass range is motivated by \cite{2019ApJ...883L..18G} and \cite{2020ARA&A..58..257G}. The source-frame loss-cone rate density for class A is
\begin{align}
{\cal R}_{\rm lc}^{(\rm A)} &\simeq
\frac{n_{\rm SMBH}}{10^{-2}\,{\rm Mpc^{-3}}}
\sum_{k=1}^{5}\sum_{j=1}^{3}
w_j\frac{\Gamma_{\rm lc}^{\rm (A)}(t_k|M_{{\rm SMBH},j})}{10^{-5}\,{\rm yr^{-1}}}
\nonumber\\&\times20\,{\rm Gpc^{-3}\,yr^{-1}},
\end{align}
including both captures and inspirals, and assuming a uniform time prior over which index $k$ runs.

The observer-frame rate follows from integrating over comoving volume and applying
cosmological time dilation \citep[Eq.~(27)]{Kritos:2025aqo}. Assuming ${\cal R}_{\rm lc}^{(\rm A)}$
is constant over $0\le z\le6$,
\begin{align}
R_{\rm lc}^{\rm (A)} &\simeq
\frac{{\cal R}_{\rm lc}^{\rm (A)}}{20\,{\rm Gpc^{-3}\,yr^{-1}}}
\frac{\int_{z=0}^{z=6}(1+z)^{-1}dV_{\rm com}(z)}{6.9\times10^{11}\,{\rm Mpc^3}}
\nonumber\\&\times1.4\times10^4\,{\rm yr^{-1}},
\end{align}
scaling linearly with $n_{\rm SMBH}$. Rates are summarized in
Table~\ref{tab:rates}.
\begin{table}[h]
    \centering
    \begin{tabular}{c c c c c}
    \hline
         A & ${\cal R}_{\rm capture}$ & $R_{\rm capture}$ & ${\cal R}_{\rm inspiral}$ & $R_{\rm inspiral}$ \\
           &  $\rm (Gpc^{-3}\,yr^{-1})$ & $\rm (yr^{-1})$ &  $\rm (Gpc^{-3}\,yr^{-1})$ & $\rm (yr^{-1})$ \\
    \hline
         MS & $5.4\times10^3$ & $3.8\times10^6$ & $4.7\times10^{-6}$ & $4.4\times10^{-2}$ \\
         giant & $1.5\times10^2$ & $1.1\times10^5$ & $1.6\times10^{-7}$ & $3.3\times10^{-3}$ \\
         WD & $8.6\times10^1$ & $6.0\times10^4$ & $4.9\times10^{-8}$ & $1.5\times10^{-3}$ \\
         NS & $1.5\times10^{1}$ & $1.0\times10^4$ & $1.1\times10^{-10}$ & $9.9\times10^{-7}$ \\
         BH & $2.2\times10^{1}$ & $1.5\times10^4$ & $2.8\times10^{-7}$ & $1.9\times10^{-2}$ \\
    \hline
    \end{tabular}
    \caption{Capture and inspiral loss-cone density and volume-integrated rates over $0\le z\le6$ for different stellar classes.}
    \label{tab:rates}
\end{table}

The total SMBH mass growth over $z\in[0,6]$ is
\begin{align}
    &\approx {n_{\rm SMBH}\over10^{-2}\,\rm Mpc^{-3}}\sum_{j=1}^3 w_j{\int_0^{10\,\rm Gyr}dt\,\dot{M}_{{\rm SMBH},j}(t) \over1.8\times10^6\,M_\odot}\nonumber\\&\times1.8\cdot10^{13}\,M_\odot\,\rm Gpc^{-3}
\end{align}
which is on the same order as So{\l}tan's estimated mass density~\citep[$8\times10^{13}\,M_\odot\,\rm Gpc^{-3}$;][]{1982MNRAS.200..115S}, implying gas accretion dominates SMBH growth but loss-cone effects contribute at the $\sim25\%$ level.

We find an MS TDE rate of $\approx5\times10^3\,{\rm Gpc^{-3}\,yr^{-1}}$ at $4\le z\le6$, with only a
small detectable fraction (e.g., UVEX; \citealt{Kulkarni:2021tit}) while inspirals are
negligible in all classes. For comparison, the local TDE rate density is
$\sim15$--$200\,{\rm Gpc^{-3}\,yr^{-1}}$~\citep{vanVelzen:2014dna}, below theoretical
$z\sim0$ expectations~\citep{Stone:2014wxa}, while recent candidates at $z>1$~\citep{Gu:2024nzu} imply rates of $\sim20$--$80\,{\rm Gpc^{-3}\,yr^{-1}}$ at $z\sim1$~\citep{Graham:2025tla}. In addition, ${\cal R}_{\rm MS}:{\cal R}_{\rm giant}\sim30$. Finally, stellar remnants capture at intrinsic rates of hundreds per year, and about two BH-EMRIs every century, producing GW sources, which we discuss next.

\section{Gravitational-wave background}
\label{sec:Gravitational-wave_background}

\begin{figure}
    \centering
    \includegraphics[width=0.49\textwidth]{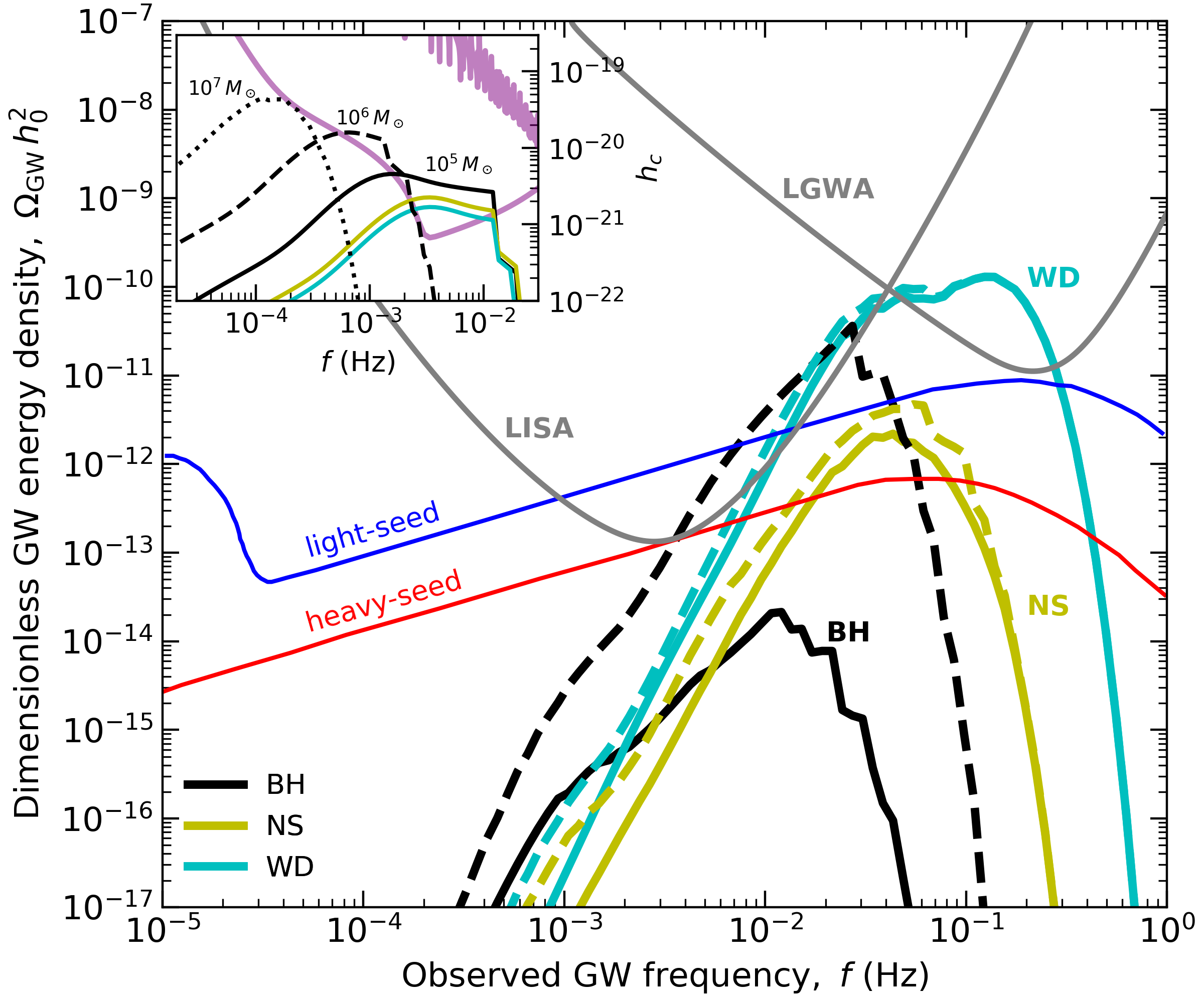}
    \caption{Total dimensionless gravitational-wave (GW) energy density from BH (black), NS (yellow), and WD (cyan) captures within $z<6$ as a function of the observed GW frequency. The comoving SMBH number density is assumed constant at $10^{-2}\,\rm Mpc^{-3}$. Gray curves indicate the PLS of LISA and LGWA with an $\rm SNR=10$ and $T_{\rm obs}=4\,\rm yr$. Also shown are the GW backgrounds from mergers of massive BHs under the light- and heavy-seed scenarios~\citep{Caliskan:2025esi}. The inset shows the characteristic strain for a few $z\sim5$ captures.}
    \label{fig:sgwb}
\end{figure}

We compute the characteristic GW strain of each capture, $h_c(f)$, at the observer-frame
frequency $f$, using Eq.~(17) of \citet{Bonetti:2020jku}, including the first 100
harmonics. Merger–ringdown and spin effects are neglected. Example individual
sources at $z\sim5$ are shown in the inset of Fig.~\ref{fig:sgwb} together with the
sensitivities of LGWA~\citep{LGWA:2020mma} and LISA~\citep{LISA:2017pwj}. Captures originate from highly eccentric
($e\simeq1^{-}$) orbits that rapidly circularize and merge within $\simeq28\,{\rm min}\,M_{\rm SMBH}/(10^6\,M_\odot)({10\,\rm km\,s^{-1}}/\sigma)^3$~\citep{OLeary:2008myb}. We find that $\simeq54\%$ ($\simeq1.3\%$), $\simeq7.9\%$ ($\simeq0.6\%$), and $\simeq3.3\%$ ($\simeq0.5\%$) of all BH, NS, and WD captures, respectively, have signal-to-noise ratio (SNR) $>10$ in LISA (LGWA).

The cumulative contribution of captures produces a stochastic GW background (SGWB),
whose dimensional energy density over an observation time $T_{\rm obs}$ we approximate as~\citep[Eq.~(3.83)]{Belgacem:2024ohp}
\begin{align}
    \Omega_{\rm lc}^{\rm (A)}(f)h_0^2 &= {1\,\rm yr\over T_{\rm obs}} \left({f\over10^{-2}\,\rm Hz}\right)^3 \sum_{i=1}^{\tilde{\cal N}_{\rm A}} \sum_{j=1}^{3}w_j \left[{h_{c,ij}(f)\over10^{-20}}\right]^2\nonumber\\&\times 2\cdot10^{-13}\,{\tilde{\cal N}_{\rm A}\over 10^4},
\end{align}
${\rm A}=\{\rm WD,\,NS,\,BH\}$ and $h_0$ is the reduced Hubble's constant. The sum is over undetectable $\tilde{\cal N}_{\rm A}$ captures with $\rm SNR<10$, by convention.

The resulting SGWBs are shown in Fig.~\ref{fig:sgwb} (thick lines). The WD
capture background peaks at dHz and are observable with an $\rm SNR>10$ when compared with the corresponding LGWA Power-Law integrated Sensitivity~\citep[PLS;][]{Ajith:2024mie} and LISA PLS~\citep{Schmitz:2020rag} for $T_{\rm obs}=4\,\rm yr$. For comparison, we also show the background from massive BH mergers under two seeding scenarios~\citep{Caliskan:2025esi}. While the BH and NS backgrounds have $\rm SNR<10$ within $\rm 4\, yr$, there are still thousands of captures within $z<6$ with LISA. Finally, the SGWB from inspirals is undetectable.

\section{Formation of the black hole star}
\label{sec:Formation_of_the_black_hole_star}

We adopt a BH–star framework in which the SMBH is embedded in a dense
cocoon of gas and stars. This model
satisfies spectral constraints on LRDs and enables rapid SMBH growth via
bursty, super-Eddington gas accretion~\citep{2025arXiv251107515K,Kritos:2025aqo}. The buildup
of dense gas cocoons produces extreme Balmer breaks, Balmer absorption, and
H$\beta$ emission,
and resembles the phenomenon of
BH-stars~\citep{2025A&A...701A.168D,Naidu:2025rpo,Begelman:2025upi,2026ApJ...998..124N,2026Natur.649..574R}.

Feedback-regulated growth yields universal scalings: momentum conservation gives
$M_{\rm BH}\propto\sigma^4$,
\begin{equation}
f_{\rm Edd}\left(\frac{4\pi G M_{\rm SMBH} c}{\kappa_{\rm es}}\right)
=\frac{\epsilon\sigma^4 v_s}{G},
\end{equation}
while energy conservation implies $M_{\rm SMBH}\propto\sigma^5$. The SMBH radius of
influence,
$r_{\rm infl}\sim10$--$30\,\rm pc$, defines the Bondi radius.

Accretion-disk sizes inferred from reverberation mapping are typically parsec scale.
The Eddington feedback region is set by requiring a unit Thomson optical depth,
\begin{equation}
({n_e\sigma_T})^{-1}
= m_p M_g^2 G^3 \sigma^{-6}\sigma_T^{-1}
\simeq (M_g/M_\odot)^2\,\sigma_{200}^{-6},
\end{equation}
using $\rho_g=\sigma^6 G^{-3} M_g^{-2}$ and stellar-mass form factors
\citep{1995MNRAS.275..498D}. This defines a dense central core at radii
$\sim1$--$10\,{\rm au}$, consistent with CLOUDY models of inverted LRD spectra
and inferred gas densities $n_{\rm H}\sim10^8\,{\rm cm^{-3}}$~\citep{2025ApJ...989L...7T}.

Observations and theory favor Comptonization-regulated feedback~\citep{2022A&A...666A..17G}: momentum-driven regulation dominates at high redshift, while
energy-driven feedback with momentum boosting becomes important at low redshift,
enhancing early star formation and later driving massive outflows~\citep{Silk:2024rsf,Costa:2014lga}. 

\section{Conclusions}
\label{sec:Conclusions}

We modeled stellar feeding onto SMBHs in NSCs, including TDEs and compact-object captures. Stellar plunges contribute only a sizable fraction to SMBH mass growth; however, gas accretion still dominates at all redshifts.

Main-sequence TDEs dominate the electromagnetic output. Disruptions of massive
($\sim30$–$150\,M_\odot$) stars by $10^{6}$–$10^{7}\,M_\odot$ SMBHs naturally reproduce the luminosities and timescales of ENTs, while giant-star TDEs are negligible. We predict an intrinsic MS TDE rate of
$\sim5\times10^{3}\,\rm Gpc^{-3}\,yr^{-1}$ at $4\le z\le6$. A fraction of these are expected to be observed with the LSST~\citep{Karmen:2026ixp}.

Compact-object captures produce thousands of detectable LISA sources within $z<6$ and generate a stochastic GW background, dominated by WD captures at dHz frequencies.

Stellar feeding is therefore an inevitable byproduct of dense, rapidly growing high-redshift nuclei; insufficient for SMBH growth, but observable through extreme TDEs and GWs, providing a direct probe of BH–star systems and early SMBH assembly.

Our data and code are publicly available at~\url{https://zenodo.org/records/18471816}.

\begin{acknowledgments}
We thank Mesut Çalışkan, Marco Chiaberge, Muryel Guolo, Francesco Iacovelli, Mitchell Karmen, Julian Krolik, Colin Norman, David Pereñiguez, and the anonymous referee for discussions and comments.
K.K. is supported by NSF grant Nos.~AST-2307146, PHY-2513337, PHY-090003, and PHY-20043, by NASA grant No.~21-ATP21-0010, by John Templeton Foundation grant No.~62840, by the Simons Foundation [MPS-SIP-00001698, E.B.], by the Simons Foundation International, by Italian Ministry of Foreign Affairs and International Cooperation grant No.~PGR01167, and by the Onassis Foundation Scholarship (ID: F ZT 041-1/2023-2024).
\end{acknowledgments}

\appendix
\section*{Extended tables of loss-cone rates}
\label{sec:Extended_tables_of_loss-cone_rates}

This Appendix presents capture loss-cone rates and associated uncertainties for each stellar type at the five different epochs onto a $10^{5}$ (Table~\ref{tab:lc-rates-5}), $10^{6}$ (Table~\ref{tab:lc-rates-6}), and $10^{7}\,M_\odot$ (Table~\ref{tab:lc-rates-7}) SMBH.

\begin{table}[]
    {\setlength{\tabcolsep}{9.1pt}
    \begin{tabularx}{\textwidth}{c c c c c c}
    \hline
        stellar class & & & $t\ (\rm Myr)$ & & \\
          & 1 & 10 & 100 & 1000 & 10,000 \\
    \hline
         MS & $(6.7\pm1.2)\times10^{-4}$ & $(7\pm2)\times10^{-4}$ & $(5.3\pm1.0)\times10^{-4}$ & $(4.3\pm0.5)\times10^{-4}$ & $(3.4\pm0.6)\times10^{-4}$ \\
         giant & 0 & $(2.0\pm1.3)\times10^{-5}$ & $(1.8\pm0.3)\times10^{-5}$ & $(3.1\pm1.2)\times10^{-5}$ & $(6.7\pm1.3)\times10^{-6}$ \\
         WD & 0 & 0 & $(8\pm4)\times10^{-7}$ & $(1.3\pm1.3)\times10^{-5}$ & $(3\pm2)\times10^{-5}$ \\
         NS & 0 & 0 & $(2.2\pm0.7)\times10^{-6}$ & $(2.3\pm0.8)\times10^{-6}$ & $(3\pm2)\times10^{-6}$ \\
         BH & 0 & $(5.2\pm0.7)\times10^{-6}$ & $(1.9\pm1.8)\times10^{-6}$ & $(1.9\pm1.8)\times10^{-6}$ & $(1.9\pm1.8)\times10^{-6}$ \\
    \hline
    \end{tabularx}
    }
    \caption{Capture loss-cone rates onto $M_{\rm SMBH}=10^{5}\,M_\odot$ in $\rm yr^{-1}$ at evolution time $t$ for different stellar types. Rates of naked cores are zero. Error bars are standard deviations estimated over 10 realizations. Total number of stellar objects within the influence radius is $N_\star=301,550$. Abbreviations: MS (main sequence); WD (white dwarf); NS (neutron star); BH (black hole).}
    \label{tab:lc-rates-5}
\end{table}

\begin{table}[]
    {\setlength{\tabcolsep}{10.5pt}
    \begin{tabularx}{\textwidth}{c c c c c c}
    \hline
        stellar class & & & $t\ (\rm Myr)$ & & \\
          & 1 & 10 & 100 & 1000 & 10,000 \\
    \hline
         MS & $(8\pm3)\times10^{-4}$ & $(6\pm4)\times10^{-4}$ & $(3.8\pm0.2)\times10^{-4}$ & $(3.8\pm0.2)\times10^{-4}$ & $(2.4\pm0.2)\times10^{-4}$ \\
         giant & 0 & $(0.5\pm0.5)\times10^{-5}$ & $(1.4\pm0.2)\times10^{-5}$ & $(2.0\pm1.3)\times10^{-5}$ & $(5.3\pm1.0)\times10^{-6}$ \\
         WD & 0 & 0 & $(1.2\pm0.2)\times10^{-6}$ & $(1.0\pm0.3)\times10^{-5}$ & $(2.6\pm0.6)\times10^{-5}$ \\
         NS & 0 & 0 & $(4\pm2)\times10^{-6}$ & $(3\pm2)\times10^{-6}$ & $(3\pm2)\times10^{-6}$ \\
         BH & 0 & $(9\pm4)\times10^{-6}$ & $(1.1\pm0.7)\times10^{-5}$ & $(1.1\pm0.7)\times10^{-5}$ & $(1.1\pm0.7)\times10^{-5}$ \\
    \hline
    \end{tabularx}
    }
    \caption{Same as Table~\ref{tab:lc-rates-5} with $M_{\rm SMBH}=10^{6}\,M_\odot$. Total number of stellar objects within the influence radius is $N_\star=3,015,502$.}
    \label{tab:lc-rates-6}
\end{table}

\begin{table}[]
    {\setlength{\tabcolsep}{7.5pt}
    \begin{tabularx}{\textwidth}{c c c c c c}
    \hline
        stellar class & & & $t\ (\rm Myr)$ & & \\
          & 1 & 10 & 100 & 1000 & 10,000 \\
    \hline
         MS & $(4.9\pm0.2)\times10^{-4}$ & $(4.3\pm0.4)\times10^{-4}$ & $(3.5\pm0.2)\times10^{-4}$ & $(3.0\pm0.1)\times10^{-4}$ & $(2.29\pm0.06)\times10^{-4}$ \\
         giant & 0 & $(1.0\pm0.4)\times10^{-5}$ & $(7.4\pm1.6)\times10^{-6}$ & $(1.2\pm0.1)\times10^{-5}$ & $(3.2\pm0.6)\times10^{-6}$ \\
         WD & 0 & 0 & $(1.9\pm0.2)\times10^{-6}$ & $(1.49\pm0.07)\times10^{-5}$ & $(4.2\pm0.3)\times10^{-5}$ \\
         NS & 0 & 0 & $(3.8\pm0.2)\times10^{-6}$ & $(3.8\pm0.2)\times10^{-6}$ & $(3.9\pm0.4)\times10^{-6}$ \\
         BH & 0 & $(1.7\pm0.6)\times10^{-5}$ & $(1.9\pm0.7)\times10^{-5}$ & $(1.8\pm0.5)\times10^{-5}$ & $(2.0\pm0.3)\times10^{-5}$ \\
    \hline
    \end{tabularx}
    }
    \caption{Same as Table~\ref{tab:lc-rates-5} with $M_{\rm SMBH}=10^{7}\,M_\odot$. Total number of stellar objects within the influence radius is $N_\star=30,155,026$.}
    \label{tab:lc-rates-7}
\end{table}

\bibliographystyle{aasjournal}
\bibliography{High-z-TDE-Rates}

\end{document}